\documentclass[showpacs]{revtex4}
\usepackage{epsfig}
\def\beq{\begin{equation}}
\def\enq{\end{equation}}
\def\beqa{\begin{eqnarray}}
\def\enqa{\end{eqnarray}}

\def\MeV{\nobreak\,\mbox{MeV}}
\def\GeV{\nobreak\,\mbox{GeV}}

\def\qq{\lag\bar{q}q\rag}

\def\mix{\lag\bar{q}g\si.Gq\rag}

\def\Gd{\lag g^2G^2\rag}
\def\G3{\lag g^3G^3\rag}

\def\rh{\rho}
\def\si{\sigma}

\def\al{\alpha}
\def\be{\beta}
\def\alma{\alpha_{max}}
\def\almi{\alpha_{min}}
\def\bemi{\beta_{min}}
\def\lb{\label}
\def\nn{\nonumber}

\newcommand{\rag}{\rangle}
\newcommand{\lag}{\langle}

\begin{document}

\title{\sc
QCD sum rules study of $QQ-\bar{u}\bar{d}$ mesons
}
\author{Fernando S. Navarra}
\email{navarra@if.usp.br}
\affiliation{Instituto de F\'{\i}sica, Universidade de S\~{a}o Paulo,
C.P. 66318, 05389-970 S\~{a}o Paulo, SP, Brazil}
\author{Marina Nielsen}
\email{mnielsen@if.usp.br}
\affiliation{Instituto de F\'{\i}sica, Universidade de S\~{a}o Paulo,
C.P. 66318, 05389-970 S\~{a}o Paulo, SP, Brazil}
\author{Su Houng Lee}
\email{suhoung@phya.yonsei.ac.kr}
\affiliation{Institute of Physics and Applied Physics, Yonsei University,
Seoul 120-749, Korea}

\begin{abstract}
We use QCD  sum rules to study the possible existence of
$QQ-\bar{u}\bar{d}$ mesons,
assumed to be a state with $J^{P}=1^{+}$. For definiteness, we work with
a current with an axial heavy diquark and a scalar light antidiquark,
at leading order in $\alpha_s$. We consider the contributions of
condensates up to dimension eight.
For the $b$-quark, we predict $M_{T_{bb}}= (10.2\pm 0.3)~{\rm GeV}$,
which is below the $\bar{B}\bar{B}^*$ threshold. For the $c$-quark,
we predict $M_{T_{cc}}= (4.0\pm 0.2)~{\rm GeV}$, in agreement with quark
model predictions.
\end{abstract}

\maketitle


The general idea of possible stable heavy tetraquarks has been first
suggested by Jaffe \cite{jaf}. The case of a tetraquark $QQ\bar{u}\bar{d}$
with quantum numbers $I=0, ~J=1$ and $P=+1$ which, following ref.\cite{ros},
we call $T_{QQ}$, is especially interesting. As already noted previously
\cite{ros,zsgr}, the $T_{bb}$ and $T_{cc}$ states cannot decay strongly
or electromagnetically into two $\bar{B}$ or two $D$ mesons in the
$S$ wave due to angular momentum conservation nor in $P$ wave due to
parity conservation. If their masses are below the $\bar{B}\bar{B^*}$
and $DD^*$ thresholds, these decays are also forbidden. Moreover, in the
large $m_Q$ limit, the light degrees of freedom cannot resolve the closely
bound $QQ$ system. This results in bound states similar to the
$\bar{\Lambda}_Q$ states, with $QQ$ playing the role of the heavy antiquark
\cite{chow}. Therefore, the stability of $\bar{\Lambda}_Q$ implies that
$QQ\bar{u}\bar{d}$ is also safe from decaying through $QQ\bar{u}\bar{d}\to
QQq+\bar{q}\bar{u}\bar{d}$ . As a result, $T_{QQ}$ is stable
with respect to strong interactions and must decay weakly.

There are some predictions for the masses of the $T_{QQ}$ states.
In ref.~\cite{zhu} the authors use a color-magnetic interaction, with
flavor symmetry breaking corrections, to study heavy tetraquarks. They
assume that the Belle resonance, $X(3872)$, is a $cq\bar{c}\bar{q}$
tetraquark, and use its mass as input to determine the mass of other
tetraquark states. They get $M_{T_{cc}}=3966~\MeV$ and $M_{T_{bb}}=
10372~\MeV$. In ref.~\cite{ros}, the authors use  one-gluon exchange
potentials and two different spatial configurations to study the
mesons $T_{cc}$ and $T_{bb}$. They get $M_{T_{cc}}=3876 - 3905~\MeV$
and $M_{T_{bb}}=10519 - 10651~\MeV$. There are also calculations using
expansion in the harmonic oscillator basis \cite{sem}, and variational 
method \cite{brst}.

In this work we use QCD sum rules (QCDSR) \cite{svz,rry,SNB}, to
study the two-point functions of the state $T_{QQ}$. There are
several reasons, why it is interesting to investigate this
channel.  First of all, having two heavy quarks, it is an explicit
exotic state.  The experimental observation would already prove
the existence of the tetraquark state without any theoretical
extrapolation.  Moreover, from a technical point of view, this
means that there are no contributions from the disconnected
diagrams, which are technically very difficulty to estimate in QCD
sum rules or in lattice gauge theory calculation.

In previous calculations, the QCDSR approach was used to study
the light scalar mesons \cite{LATORRE,SN4,sca,koch,zhang} the
$D_{sJ}^+(2317)$ meson \cite{pec,OTHERA} and the $X(3872)$ meson
\cite{x3872}, considered as four-quark states
and a good agreement with the experimental masses was obtained.
However, the tests were not decisive as the usual quark--antiquark
assignments also provide predictions consistent with data
\cite{SN4,SNHEAVY,SNB,OTHER}.


Considering $T_{QQ}$ as an axial diquark-antidiquark state, a possible
current describing such state is given by:
\beq
j_\mu=i[Q_a^TC\gamma_\mu Q_b][\bar{u}_a\gamma_5 C\bar{d}_b^T]\;,
\label{field}
\enq
where $a,~b$ are color indices, $C$ is the charge conjugation
matrix and $Q$ denotes the heavy quark.

In general, one should consider all possible combinations of
different $1^{+}$ four-quark operators, as was done in
\cite{chinois}  for the $0^{++}$ light mesons. However, the
current in Eq.(\ref{field}) well represents the most attractive
configuration expected with two heavy quarks.  This is so because
the most attractive light antidiquark is expected to be the in the
color triplet, flavor anti-symmetric and spin 0
channel \cite{Jaffe03,Shuryak03,Schafer93}.  This is also expected
quite naturally from the color magnetic interaction, which can be
phenomenologically parameterized as,
\begin{eqnarray}
V_{ij}=-\frac{C}{m_i m_j} \lambda_i \cdot \lambda_j \sigma_i \cdot
\sigma_j. \label{color-magnetic}
\end{eqnarray}
Here, $m,\lambda, \sigma$ are the mass, color and spin of the
constituent quark $i,j$.  Eq.(\ref{color-magnetic}) favors the
anti-diquark to be in the color triplet and spin 0 channel. The
flavor anti-symmetric condition then follows from requiring
anti-symmetric wave function of the  anti-diquark. Similarly,
since the anti-diquark is in the color triplet state the remaining
$QQ$ should be in the color anti-triplet spin 1 state. Although
the spin 1 configuration is repulsive, its strength is much
smaller than that for the light diquark due to the heavy charm
quark mass.  Therefore a constituent quark picture for $T_{QQ}$
would be a light anti-diquark in color triplet, flavor  anti-symmetric and
spin 0 ($\epsilon_{abc}[\bar{u}_b\gamma_5 C\bar{d}_c^T]$) combined
with a heavy diquark of spin 1 ($\epsilon_{aef}[Q_e^TC\gamma_\mu
Q_f]$).  The simplest choice for the current to have a non zero
overlap with such a $T_{QQ}$ configuration is given in
Eq.~(\ref{field}).   While a similar configuration $T_{ss}$ is
also possible \cite{Morimatsu}, we believe that the repulsion in
the strange diquark with spin 1 will be larger and hence
energetically less favorable.   As discussed above, since the
quantum number is $1^+$, the decay into $DD$ or $\bar{B}\bar{B}$
would be forbidden and the allowed decay into $DD^*$ or
$\bar{B}\bar{B}^*$ would have a smaller phase space, and the
tetraquark state might have a small width, or may even be bound.

The QCDSR is constructed from the two-point correlation function
\beq
\Pi_{\mu\nu}(q)=i\int d^4x ~e^{iq.x}\lag 0
|T[j_\mu(x)j^\dagger_\nu(0)]
|0\rag=-\Pi_1(q^2)(g_{\mu\nu}-{q_\mu q_\nu\over q^2})+\Pi_0(q^2){q_\mu
q_\nu\over q^2}.
\lb{2po}
\enq
Since the axial vector current is not conserved, the two functions,
$\Pi_1$ and $\Pi_0$, appearing in Eq.~(\ref{2po}) are independent and
have respectively the quantum numbers of the spin 1 and 0 mesons.

The calculation of the
phenomenological side proceeds by inserting intermediate states for
the meson $T_{QQ}$.  Parametrizing the coupling of the axial vector meson
$1^{+}$, to the current, $j_\mu$, in Eq.~(\ref{field}) in terms
of the meson decay constant $f_T$ and the meson mass $M_T$ as:
\beq\label{eq: decay}
\lag 0 |
j_\mu|T_{QQ}\rag =\sqrt{2}f_T M_T^4\epsilon_\mu~,
\enq
the phenomenological side
of Eq.~(\ref{2po}) can be written as
\beq
\Pi_{\mu\nu}^{phen}(q^2)={2f_T^2M_T^8\over
M_T^2-q^2}\left(-g_{\mu\nu}+ {q_\mu q_\nu\over M_T^2}\right)
+\cdots\;, \lb{phe} \enq
where the Lorentz structure $g_{\mu\nu}$ gets contributions only from
the $1^{+}$ state.  The dots
denote higher axial-vector resonance contributions that will be
parametrized, as usual, through the introduction of a continuum
threshold parameter $s_0$ \cite{io1}.

On the OPE side, we work at leading order in $\alpha_s$ and consider the
contributions of condensates up to dimension eight.  To keep the 
charm quark mass finite, we use the momentum-space expression for the 
charm quark propagator.  We follow ref.~\cite{shl} and calculate
the light quark part of the correlation function 
in the coordinate-space, which is then Fourier transformed to
the momentum space in $D$ dimensions.  The resulting light-quark part
is combined with the charm quark part before it is dimensionally
regularized at $D=4$.

The correlation function, $\Pi_1$, in the OPE side can be written as a
dispersion relation:
\beq
\Pi_1^{OPE}(q^2)=\int_{4m_Q^2}^\infty ds {\rho(s)\over s-q^2}\;,
\lb{ope}
\enq
where the spectral density is given by the imaginary part of the
correlation function: $\pi \rho(s)=\mbox{Im}[\Pi_1^{OPE}(s)]$.  After
making a Borel transform of both sides, and
transferring the continuum contribution to the OPE side, the sum rule
for the axial vector meson $T_{QQ}$ up to dimension-eight condensates can
be written as:
\beq 2f_T^2M_T^8e^{-M_T^2/M^2}=\int_{4m_Q^2}^{s_0}ds~
e^{-s/M^2}~\rho(s)\; +\Pi_1^{mix\qq}(M^2)\;, \lb{sr} \enq
where
\beq
\rho(s)=\rho^{pert}(s)+\rh^{\qq}(s)+\rh^{\lag G^2\rag}(s)
+\rh^{mix}(s)+\rh^{\qq^2}(s)+\rh^{mix\qq}(s)\;,
\lb{rhoeq}
\enq
with
\beqa\label{eq:pert}
&&\rho^{pert}(s)={1\over 2^{9} \pi^6}\int\limits_{\almi}^{\alma}
{d\al\over\alpha^3}
\int\limits_{\bemi}^{1-\al}{d\be\over\be^3}(1-\al-\be)
\left[(\al+\be)m_Q^2-\al\be s\right]^3
\nn\\
&&\times\left[{1+\al+\be\over4}\left((\al+\be)m_Q^2-\al\be s\right)
-m_Q^2(1-\al-\be)\right],
\nn\\
&&\rho^{\qq}(s)=0,
\nn\\
&&\rho^{\lag G^2\rag}(s)=-{\Gd\over2^{10}\pi^6}
\left\{-{1\over4}\int\limits_{\almi}^{\alma} {d\al\over\al(1-\al)}
(m_Q^2-\al(1-\al)s)^2\right.\nn\\
&+&
\int\limits_{\almi}^{\alma} {d\al\over\al}
\int\limits_{\bemi}^{1-\al}d\be\left[{(\al+\be)m_Q^2-\al\be s\over4\be}
\left((\al+\be)m_Q^2-\al\be s+2m_Q^2\right)\right.
\nn\\
&+&{m_Q^2\over3\al^2}(1-\al-\be)\left[m_Q^2(1-\al-\be)+\left((\al+\be)m_Q^2
-\al\be s\right)\left(-4-\al-\be+{3\over\be}(1-\al)\right)\right]
\nn\\
&+&\left.\left.{1\over48\al\be^2}(1-\al-\be)\left((\al+\be)m_Q^2-\al\be
s\right)^2(5-\al-\be)\right]\right\},
\nn\\
&&\rho^{mix}(s)=0,
\nn\\
&&\rho^{\qq^2}(s)={\qq^2\over 24\pi^2}s\sqrt{1-4m_Q^2/s}.
\enqa
where the integration limits are given by $\almi=({1-\sqrt{1-
4m_Q^2/s})/2}$, $\alma=({1+\sqrt{1-4m_Q^2/s})/2}$ and $\bemi={\al
m_Q^2/( s\al-m_Q^2)}$.
The contribution of dimension-six condensates $\lag g^3 G^3\rag$
is neglected, since it is assumed to be suppressed by   the loop
factor $1/16\pi^2$. We have included, for completeness, a part of
the dimension-8 condensate contributions. We should note that a
complete evaluation of these contributions require more involved
analysis including a non-trivial choice of the factorization
assumption basis \cite{BAGAN} \beqa
&&\rho^{mix\qq}(s)=-{\mix\qq\over 2^6\pi^2}\sqrt{1-4m_Q^2/s},
\nn\\
&&\Pi_1^{mix\qq}(M^2)=-{m_Q^2\mix\qq\over 2^53\pi^2}\int_0^1
d\al\,\bigg[4-{m_Q^2\over\al(1-\al) M^2}\bigg]\,\exp\!\left[{-{m_Q^2
\over\al(1-\al)M^2}}\right].
\label{dim8}
\enqa
%

In order to extract the mass $M_T$ without worrying about the value of
the decay constant $f_T$, we take the derivative of Eq.~(\ref{sr})
with respect to $1/M^2$, divide the result by Eq.~(\ref{sr}) and
obtain:
\beq
M_T^2={\int_{4 m_Q^2}^{s_0}ds ~e^{-s/M^2}~s~\rho(s)\over\int_{4
m_Q^2}^{s_0}ds ~e^{-s/M^2}~\rho(s)}\;.
\lb{m2}
\enq
This quantity has the advantage to be less sensitive to the perturbative
radiative corrections than the individual
moments. Therefore, we expect that our results obtained  to leading order
in $\alpha_s$ will be quite accurate.

In the numerical analysis of the sum rules, the values used for the
quark
masses and condensates are (see e.g.  \cite{SNB,narpdg}):
$m_c(m_c)=(1.23\pm 0.05)\,\GeV $, $m_b(m_b)=(4.24\pm 0.06)\,\GeV$,
$\lag\bar{q}q\rag=\,-(0.23\pm0.03)^3\,\GeV^3$,
$\lag\bar{q}g\si.Gq\rag=m_0^2\lag\bar{q}q\rag$ with $m_0^2=0.8\,\GeV^2$,
$\lag g^2G^2\rag=0.88~\GeV^4$.

\begin{figure}[h]
\centerline{\epsfig{figure=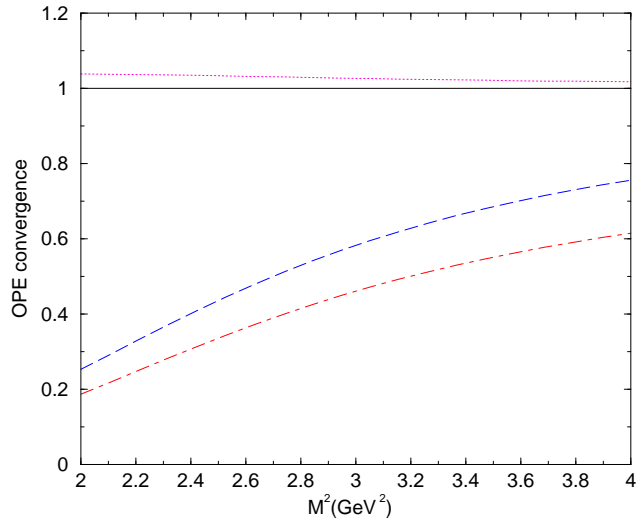,height=70mm}}
\caption{The relative OPE convergence in the region $2.0 \leq M^2 \leq
4.0~\GeV^2$ for $\sqrt{s_0} = 4.8$ GeV. We start with the perturbative
contribution divided by the total (long-dashed line) and each
subsequent line represents the addition of one extra condensate
dimension in the expansion: $+\langle g^2G^2\rangle$ (dot-dashed line),
$+\langle \bar{q}q\rangle^2$ (dotted-line), $+ m_0^2\langle \bar{q}q
\rangle^2$ (solid line).}
\label{figconvtcc}
\end{figure}

We start with the double charmed meson $T_{cc}$.
We evaluate the sum rules in the range $2.0 \leq M^2 \leq 4\GeV^2$ for
$s_0$ in the range: $4.6\leq \sqrt{s_0} \leq5.0$ GeV.

\begin{figure}[h]
\centerline{\epsfig{figure=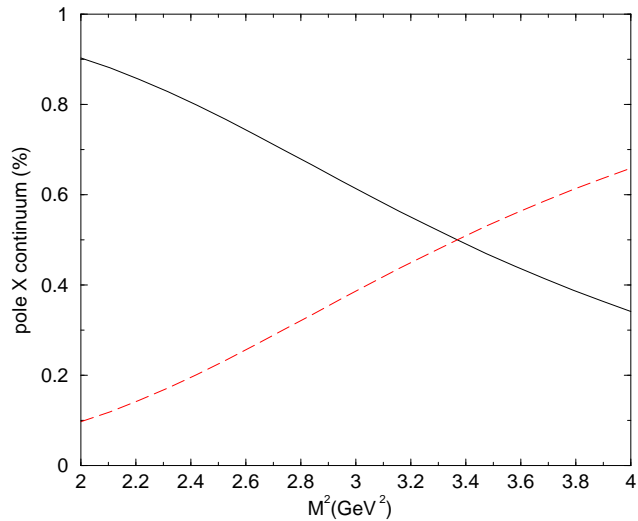,height=70mm}}
\caption{The solid line shows the relative pole contribution (the
pole contribution divided by the total, pole plus continuum,
contribution) and the dashed line shows the relative continuum
contribution for $\sqrt{s_0}=4.8~\GeV$.}
\label{figpvc}
\end{figure}

\begin{figure}[h]
\centerline{\epsfig{figure=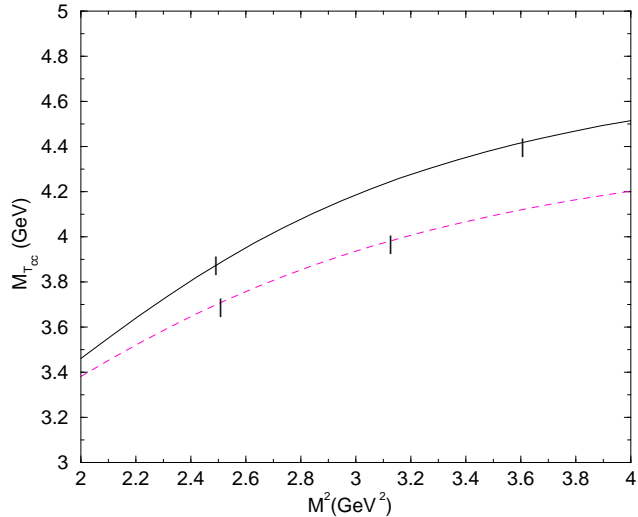,height=70mm}}
\caption{The $T_{cc}$ meson mass as a function of the sum rule parameter
($M^2$) for different values of the continuum threshold: $\sqrt{s_0} =
4.6$ GeV (dotted line) and $\sqrt{s_0} = 5.0$ GeV (solid line).  The
bars indicate the region allowed for the sum rules: the lower limit
(cut below 2.5 GeV$^2$) is given by OPE convergence requirement and the
upper limit by the dominance of the QCD pole contribution.}
\label{figmx}
\end{figure}

\begin{figure}[h]
\centerline{\epsfig{figure=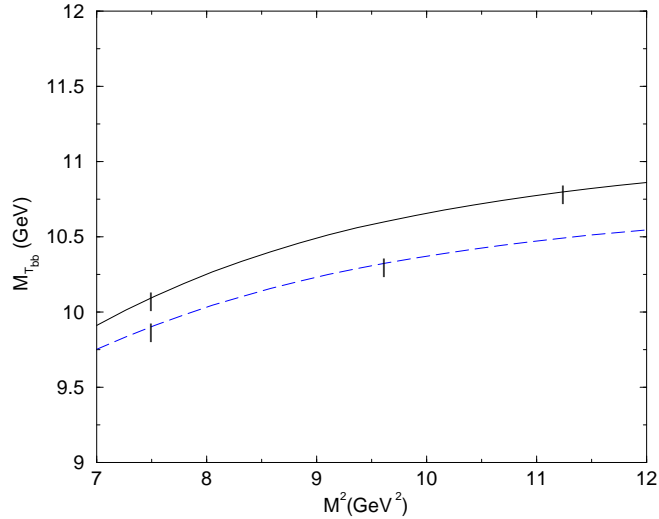,height=70mm}}
\caption{The $T_{bb}$ meson mass as a function of the sum rule parameter
($M^2$) for different values of the continuum threshold:
$\sqrt{s_0}=11.3$ GeV (dashed line), $\sqrt{s_0}=11.7$ GeV (solid
line).  The bars delimit the region allowed for the  sum rules.}
\label{figmxb}
\end{figure}

Comparing the relative contribution of each term in
Eqs.~(\ref{eq:pert}) to (\ref{dim8}),
to the right hand side of Eq.~(\ref{sr}) we obtain a quite good OPE
convergence (the perturbative contribution is at least 50\% of the total)
for $M^2 > 2.5$ GeV$^2$, as can be seen in
Fig.~\ref{figconvtcc}.  This analysis allows us to determine the lower
limit constraint for $M^2$ in the sum rules window.  This figure also
shows that, although there is a change of sign between
dimension-six and dimension-eight condensates contributions, the
contribution of the latter is very small, where, we have assumed, in
Fig.~\ref{figconvtcc} to  Fig.~\ref{figmxb}, the validity of
the vacuum saturation for these condensates. The relatively small
contribution of the dimension-eight condensates may justify  the validity
of our approximation, unlike in the case of the 5-quark current
correlator, as noticed in \cite{oganes}.

We get an upper limit constraint for $M^2$ by imposing the rigorous
constraint that the QCD continuum contribution should be smaller than the
pole contribution.
The maximum value of $M^2$ for which this constraint is satisfied
depends on the value of $s_0$.  The comparison between pole and
continuum contributions for $\sqrt{s_0} = 4.8$ GeV is shown in
Fig.~\ref{figpvc}. The same analysis for the other values of the continuum
threshold gives $M^2 \leq 3.1$  GeV$^2$ for $\sqrt{s_0} = 4.6~\GeV$ and
$M^2 \leq 3.6$  GeV$^2$ for $\sqrt{s_0} = 5.0~\GeV$.

In Fig.~\ref{figmx}, we show the $T_{cc}$ meson mass obtained from
Eq.~(\ref{m2}), in the relevant sum rules window, with the upper and
lower validity limits indicated.  From Fig.~\ref{figmx} we see that
the results are reasonably stable as a function of $M^2$.
In our numerical analysis, we shall then consider the range of $M^2$ values
from 2.5 $\GeV^2$ until the one allowed by the sum rule window criteria as
can be deduced from Fig.~\ref{figmx} for each value of $s_0$.

We found that our results are not very sensitive to the value of the
charm quark mass, neither to the value of the condensates. The most
important source of uncertainty is the value
of the continuum threshod and the Borel interval. Using the QCD parameters
given above, the QCDSR predictions for the $T_{cc}$ mesons mass is:
\beq
M_{T_{cc}} = (4.0\pm0.2)~\GeV,
\enq
in a very good agreement with the predictions in refs.~\cite{ros} and
\cite{zhu}.

One can also evaluate the decay constant, defined in Eq.~(\ref{eq: decay}),
to leading order in $\alpha_s$:
\beq\label{fx}
f_{T_{cc}}=(5.95\pm 0.65)\times 10^{-5}~{\rm  GeV}~,
\enq
which can be more affected by  radiative corrections than $M_{T_{cc}}$.

In the case of the double-beauty meson $T_{bb}$,
using consistently the perturbative $\overline{MS}$-mass $m_b(m_b)=(4.24
\pm0.6)~\GeV$,
and the continuum threshold in the range $11.3\leq\sqrt{s_0}\leq11.7~\GeV$,
we find a good OPE convergence for $M^2>7.5~\GeV^2$.
We also find that the pole contribution is bigger than the continuum
contribution for $M^2<9.6~\GeV^2$ for $\sqrt{s_0}<11.3~\GeV$, and
for $M^2<11.2~\GeV^2$ for $\sqrt{s_0}<11.7~\GeV$.

In Fig.~\ref{figmxb} we show the $T_{bb}$ meson mass obtained from
Eq.~(\ref{m2}), in the relevant sum rules window, with the upper and
lower validity limits indicated.
From Fig.~\ref{figmxb} we see that the results are very stable as a
function of $M^2$ in the allowed region. Taking into account the
variation of $M^2$ and varying $s_0$ and $m_b$ in the regions indicated
above, we arrive at the prediction:
\beq
\lb{massXb}
M_{T_{bb}}=  (10.2\pm0.3)~\GeV~,
\enq
also in a very good agreement with the results in refs.~\cite{ros},
\cite{zhu} and \cite{brst}.
For completeness, we predict the corresponding value of the decay
constant to leading order in $\alpha_s$:
\beq\label{fxb}
f_{T_{bb}}= (10.4\pm 2.8)\times 10^{-6} ~{\rm GeV}~.
\enq


We have presented a QCDSR analysis of the two-point
functions of the double heavy-quark axial meson, $T_{QQ}$, considered as a
four quark state.  We find that the sum rules results for the masses of
$T_{cc}$ and $T_{bb}$ are compatible with the results in refs.~\cite{ros}
and \cite{zhu}. An improvement of this result needs an accurate
determination of running masses $m_c$ and $m_b$ of the
${\overline{MS}}$-scheme and the inclusion of radiative corrections.

Our results show that while the $T_{cc}$ mass is bigger than the
$D^*D$ threshold at about 3.875 GeV, the $T_{bb}$ mass is
appreciably below the $\bar{B}^*\bar{B}$ threshold at about
$10.6\;$GeV. Therefore, our results indicate that the  $T_{bb}$
meson should be stable with respect to strong interactions and
must decay weakly.  Our result also confirms the naive expectation
that the exotic states with heavy quarks tend to be more stable
than the corresponding light states\cite{Lee05}.

We present in Eqs. (\ref{fx}), and (\ref{fxb})
predictions for the decay constants  of the $T_{cc}$ and $T_{bb}$.

Different choices of the four-quark operators have been systematically
presented for the $0^{++}$ light mesons in  \cite{chinois}. Though some
combinations can provide a faster convergence of the OPE, we do not
expect that the choice of the operators
will affect much our results, where, in our analysis, the OPE has a good
convergence.

\section*{Acknowledgements}
{This work has been partly supported by FAPESP and CNPq-Brazil,
and by the Korea Research Foundation KRF-2006-C00011.}


\end{document}